\documentclass{article}

\usepackage{arxiv}

\usepackage[utf8]{inputenc} 
\usepackage[T1]{fontenc}    
\usepackage{hyperref}       
\usepackage{url}            
\usepackage{booktabs}       
\usepackage{amsfonts}       
\usepackage{nicefrac}       
\usepackage{microtype}      
\usepackage{lipsum}
\usepackage{graphicx}
\usepackage{amsmath}
\usepackage{amssymb}
\usepackage{float}
\usepackage{tikz}
\graphicspath{ {./images/} }

\title{Boundary‑Informed Method of Lines for Physics‑Informed Neural Networks\thanks{This paper was accepted to NYSDS 2025 and will appear in the SIAM Proceedings.}}

\author{
 Maximilian Cederholm \\
  Department of Physics\\
  Stony Brook University\\
   \And
Haochun Wang \\
   Department of Applied Mathematics and Statistics\\
   Stony Brook University\\
   \And
 Siyao Wang \\
  Department of Statistics\\
  University of California, Davis\\
   \AND
   Ruichen Xu \\
   Department of Applied Mathematics and Statistics\\
   Stony Brook University\\
  \texttt{ruichen.xu@stonybrook.edu} \\
  \And
  Yuefan Deng \\
  Department of Applied Mathematics and Statistics\\
  Stony Brook University\\
  \texttt{yuefan.deng@stonybrook.edu} \\
}

\begin{document}
\maketitle
\begin{abstract}
  \label{sec:abstract}
  We propose a hybrid solver that fuses the dimensionality‑reduction strengths of the Method of Lines (MOL) with the flexibility of Physics‑Informed Neural Networks (PINNs). Instead of approximating spatial derivatives with fixed finite‑difference stencils—whose truncation errors force extremely fine meshes—our method trains a neural network to represent the initial spatial profile and then employs automatic differentiation to obtain spectrally accurate gradients at arbitrary nodes. These high‑fidelity derivatives define the right‑hand side of the MOL‑generated ordinary‑differential system, and time integration is replaced with a secondary temporal PINN while spatial accuracy is retained without mesh refinement. The resulting “boundary‑informed MOL‑PINN” matches or surpasses conventional MOL in accuracy using an order of magnitude fewer collocation points, thereby shrinking memory footprints, lessening dependence on large data sets, and increasing complexity robustness. Because it relies only on automatic differentiation and standard optimizers, the framework extends naturally to linear and nonlinear PDEs in any spatial dimension.
\end{abstract}


\section{Introduction}
\label{sec:intro}

Data–driven surrogates that respect physical laws are rapidly reshaping the numerical solution of differential equations.  A leading paradigm is the \emph{physics-informed neural network} (PINN), in which the governing equations appear as soft constraints in the loss function so that the network is trained not only on observational data but also on residuals of the underlying operators \cite{karniadakis2021physics}.  By embedding the residual of a partial differential equation (PDE) directly into stochastic gradient descent, PINNs circumvent the need for dense labeled data sets and can, in principle, generalize across regimes that were never explicitly observed.  However, in more than one spatial dimension the residual must be evaluated at a large number of collocation points distributed throughout the spatio-temporal domain.  Because every additional derivative of the PDE introduces a new set of boundary conditions, the effective hypothesis space grows combinatorially, and the optimizer must navigate an increasingly flat, high-dimensional loss landscape.  The upshot is that PINNs sometimes struggle to converge or require prohibitively many collocation points when confronted with stiff, multi-scale, or high-order PDEs.

One classical strategy for taming dimensionality is the \emph{Method of Lines} (MOL)
\cite{schiesser1991method}: discretize space, treat time as continuous, and solve the resulting system of ordinary differential equations (ODEs) with robust ODE integrators. Unfortunately, the spatial derivatives in standard MOL are approximated via finite-difference (FD) stencils whose truncation error scales algebraically with the mesh width; achieving high fidelity demands finely resolved grids that inflate memory consumption and computational cost \cite{leveque2007finite}. In addition, FD operators are tied to regular lattices and lose accuracy or even stability on irregular nodesets, limiting their compatibility with the randomly sampled collocation points favored in PINN training.

We propose to combine the dimensional-reduction power of MOL with the differentiability of PINNs while \emph{removing} finite-difference approximations altogether. Our key observation is that if the initial spatial profile is represented by a smooth neural
network, then all spatial derivatives can be obtained to machine precision by automatic differentiation (AD) \cite{baydin2018automatic}. By training a boundary-informed network solely on the initial slice \(t=0\), we obtain a differentiable surrogate whose gradients serve as spectrally accurate inputs to the MOL ODE system. Time evolution is then handled either by classical integrators or by a secondary PINN that operates purely in the temporal domain, effectively decoupling space and time. The resulting \emph{boundary-informed MOL-PINN} inherits three distinct advantages: (i) high-order spatial accuracy on arbitrary node sets without mesh refinement, (ii) a drastic reduction in the number of collocation points required during training, and (iii) the flexibility of modern AD frameworks, making the method immediately compatible with existing scientific machine-learning toolchains. These benefits are complementary to recent neural-operator advances that handle
symmetries \cite{gao2024coordinate}, enhance expressive power \cite{gao2025dynamic}, and diagnose discretization-mismatch errors \cite{gao2025discretizationinvariance}, as well as to structure-preserving Hamiltonian learning that preserves invariants \cite{wu2025kolmogorovarnoldrepresentationsymplecticlearning}. We demonstrate that the hybrid solver delivers solutions of comparable or superior accuracy to FD-based MOL while using an order of magnitude fewer residual evaluations, thereby lowering memory overhead and accelerating convergence for both linear and nonlinear PDEs in one and multiple dimensions.

\section{Preliminaries.}
\label{sec:prelim}

\subsection{Method of Lines (MOL).}
\label{sec:prelim:mol}
The core idea of MOL is to discretize \emph{all but one} independent variable.  For
a $d$‑dimensional spatial domain \(\Omega\subset\mathbb{R}^{d}\) and a
final time \(T>0\), consider a PDE written in residual form
\[
  \mathcal{R}[u](\mathbf{x},t)
  :=\frac{\partial^{n}u}{\partial t^{n}}(\mathbf{x},t)
     -\mathcal{S}[u](\mathbf{x},t)=0,
  \qquad (\mathbf{x},t)\in\Omega\times(0,T],
\]
where \(\mathcal{S}\) is the spatial operator and
\(n\) denotes the highest temporal derivative.  Imposing
initial data \(u(\mathbf{x},0)=u_0(\mathbf{x})\) with suitable
boundary conditions closes the problem. The spatial domain is replaced by 
\(\{\mathbf{x}_{i}\}_{i=1}^{N_x}\).  Defining
\(u_i(t):=u(\mathbf{x}_i,t)\) and introducing a \emph{discrete}
approximation \(\mathcal{S}_{\mathrm{disc}}\) of \(\mathcal{S}\), we obtain an ordinary differential system
$
  \mathrm{d}^{n}u_i/\mathrm{d}t^{n}
    =\mathcal{S}_{\mathrm{disc}}\bigl[\mathbf{u}(t)\bigr]_i,  i=1,\dots,N_x
$
and \(\mathbf{u}(t)=(u_1(t),\dots,u_{N_x}(t))^{\top}\).
While MOL cleanly separates
spatial and temporal discretizations, its accuracy is bottlenecked by
spatial truncation errors \cite{leveque2007finite}.

\subsection{Neural networks.}
\label{sec:prelim:nn}
A \emph{feed‑forward neural network} (FFNN) with $L$ hidden layers is a
parameterized mapping \(u_{\theta}:\mathbb{R}^{m}\to\mathbb{R}^{p}\)
defined as a composition of affine transformations and element‑wise
nonlinearities \cite{goodfellow2016deep}:
\[
  u_{\theta}(x)
  =f_{L}\circ f_{L-1}\circ\cdots\circ f_{1}(x),
  \quad
  f_{\ell}(x)=\sigma\!\bigl(W_{\ell}x+b_{\ell}\bigr),
  \ \ell=1,\dots,L,
\]
where \(W_{\ell}\in\mathbb{R}^{d_{\ell}\times d_{\ell-1}}\) and
\(b_{\ell}\in\mathbb{R}^{d_{\ell}}\) are weights and biases,
\(\sigma\) is a smooth activation function (e.g.\ \(\tanh\) or
$\sin$), and
\(\theta=\{W_{\ell},b_{\ell}\}_{\ell=1}^{L}\) denotes the collection of
trainable parameters.  Given data pairs
\(\{(x_i,u_{\mathrm{data}}(x_i))\}_{i=1}^{N}\), the network is typically
fitted by minimizing a mean‑squared error (MSE),
$\mathcal{L}_{\mathrm{MSE}}(\theta)
  =\frac{1}{N}\sum_{i=1}^{N}
   \bigl(u_{\theta}(x_i)-u_{\mathrm{data}}(x_i)\bigr)^{2},$
using stochastic gradient descent or its adaptive variants.

\subsection{Physics‑Informed Neural Networks (PINNs).}
\label{sec:prelim:pinn}
PINNs augment data loss with the governing physics by penalizing the PDE
residual at \emph{collocation points} \cite{raissi2019physics}.  Let
\(\{(\mathbf{x}_i^{\mathrm{int}},t_i^{\mathrm{int}})\}_{i=1}^{N_{\mathrm{int}}}\)
be interior points and
\(\{(\mathbf{x}_j^{\partial},t_j^{\partial})\}_{j=1}^{N_{\partial}}\) be
boundary points. A PINN approximates
\(u(\mathbf{x},t)\approx u_{\theta}(\mathbf{x},t)\) while enforcing
$
  \mathcal{R}\bigl[u_{\theta}\bigr](\mathbf{x}_i^{\mathrm{int}},t_i^{\mathrm{int}})
  \approx 0 \text{ and } 
  \mathcal{B}\bigl[u_{\theta}\bigr](\mathbf{x}_j^{\partial},t_j^{\partial})
  \approx g_j.
$
The composite loss
\[
  \mathcal{L}_{\mathrm{PINN}}(\theta)
  =\frac{1}{N_{\mathrm{int}}}\sum_{i=1}^{N_{\mathrm{int}}}
     \bigl\lVert
        \mathcal{R}[u_{\theta}]
        (\mathbf{x}_i^{\mathrm{int}},t_i^{\mathrm{int}})
     \bigr\rVert^{2}
   +\lambda_{\partial}
    \frac{1}{N_{\partial}}\sum_{j=1}^{N_{\partial}}
     \bigl\lVert
        \mathcal{B}[u_{\theta}]
        (\mathbf{x}_j^{\partial},t_j^{\partial})-g_j
     \bigr\rVert^{2}
\]
is minimized with respect to \(\theta\).  The approach is mesh‑agnostic;
collocation points can be drawn randomly or adaptively, making PINNs
attractive for irregular geometries.  Their main drawback is
\emph{dimensional curse} \cite{karniadakis2021physics}.

\subsection{From MOL to Boundary‑informed MOL‑PINN.}
\label{sec:prelim:molpinn}

To combine the dimensionality reduction capabilities of MOL with the differentiability and mesh flexibility of PINNs—while avoiding finite-difference errors—we propose the following two-stage workflow:

\paragraph{Stage I \textemdash\ Spatial surrogate learning.}
We restrict to \(t=0\) and train an FFNN to satisfy the \emph{boundary} residual
\begin{equation*}
  \mathcal{R}_{\partial}[u_{\theta_{\partial}}](\mathbf{x},0)
  =\frac{\partial^{n}u_{\theta_{\partial}}}{\partial t^{n}}(\mathbf{x},0)
   -\mathcal{S}[u_{\theta_{\partial}}](\mathbf{x},0)\approx0,  \qquad
    \{\, \theta_{\partial} \mid [u_{\theta_{\partial}}](\mathbf{x}, 0) \approx \mathcal{B}(\mathbf{x}) \,\}
\end{equation*}
together with boundary data.  Because time is frozen, this task is
equivalent to solving a $(d-1)$‑dimensional elliptic problem, which is
considerably easier than the full spatio‑temporal PDE.  Crucially,
automatic differentiation (AD) can now evaluate \(\mathcal{S}\)
\emph{exactly} at any $\mathbf{x}$, producing a spectrally accurate
discrete operator
\(\mathcal{S}_{\mathrm{AD}}\).

\paragraph{Stage II \textemdash\ Temporal evolution in reduced space.}
The AD‑derived operator is frozen, and the PDE reduces to an ODE system
\vspace{-1mm}
\[
\partial_t^{n}u_i(t)=\mathcal{S}_{\mathrm{AD},i},
   i=1,\dots,N_x, \qquad
  \langle \frac{\partial^{n}u_{\theta_{\partial}}}{\partial t^{n}}(\mathbf{x}_i,t)
   -\mathcal{S}_{AD}[u_{\theta_{\partial}}](\mathbf{x}_i) \rangle = \overset{\rightharpoonup}{\mathcal{R}} \approx 0
\]
\vspace{0mm}
which can be advanced either with classical solvers or, as in this work, 
with a compact \emph{temporal} PINN \(u_{\theta_t}(t)\) that depends 
\emph{only} on time. The number of \textbf{MOL trajectories} is indexed by \(i\), 
providing an adjustable granularity that balances accuracy and computational 
cost, analogous to the role of collocation points in traditional methods.

The resulting \emph{boundary‑informed MOL‑PINN} inherits three key
benefits:
   (I) \textbf{Spectral‑like spatial accuracy} without finite‑difference
        errors, thanks to AD on a smooth neural surrogate.
   (II) \textbf{Dimensionality reduction}
        during temporal integration, leading to faster convergence and higher accuracy
   (III) \textbf{Mesh independence}, enabling arbitrary node placement
        and straightforward extension to complex geometries and mixed
        operators.

These properties make the boundary‑informed MOL‑PINN a compelling candidate for stiff, multi‑scale, or data‑scarce PDEs, as empirically validated in Section \ref{sec:results}.


\begin{figure}[H]
  \centering
  \includegraphics[width=0.8\textwidth]{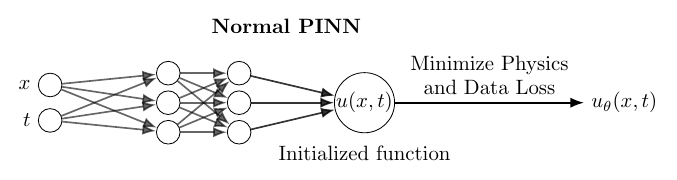}
  \vspace{5mm} 
  \includegraphics[width=0.8\textwidth]{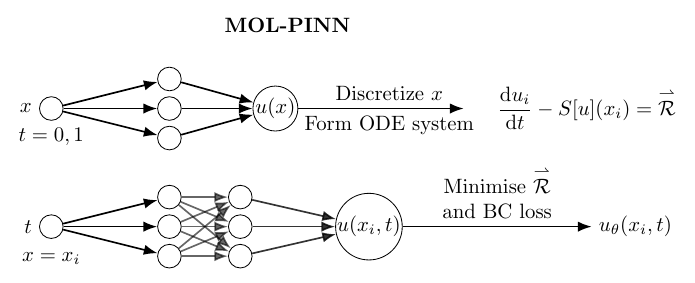}
  \caption{Visual comparison between PINN and MOL-PINN architectures.}
  \label{fig:stacked_images}
\end{figure}

\newpage
\section{Hardware and Hyper-parameters}
\label{sec:hardware}

All the relevant training hardware and hyperparameters are summarized in table 1.

\begin{table}[H]
  \centering
  \caption{Hyper-parameters and hardware.}
  \begin{tabular}{lcc}
    \hline
    \textbf{Parameter} & \textbf{Value} \\
    \hline
    Epochs & 8,000 \\
    Learning Rate (LR) & $1.0 \times 10^{-3}$ \\
    Optimizer & Adam \\
    GPU & RTX 4070 \\
    \hline
  \end{tabular}
\end{table}

\section{Results}
\label{sec:results}

All tests were run on an RTX 4070, shown in table 3.1, and all results can be found in table 3.2. Recent tests show that MOL-PINN provides an overall unusual benefit---\textbf{complexity robustness}. While a PINN's accuracy rapidly deteriorates as the PDEs increase in complexity, MOL-PINN seems to have a slower rate of deterioration. 2D nonlinear PDEs such as Burger's equation approximate at almost the same level of error as the 3D Navier Stokes. MOL-PINN properly began deteriorating at the 4D Navier Stokes, and yet it maintained a higher accuracy than PINN.

\begin{table}[H]
  \centering
  \caption{MOL-PINN MSE results \& comparison.}
  \begin{tabular}{lcccc}
  \hline
          & \textbf{PINN} & \textbf{21i} & \textbf{50i} & \textbf{100i} \\
  \hline
  Burgers         & $7.15\times10^{-3}$ & $2.40\times10^{-1}$ & $1.27\times10^{-2}$ & $2.79\times10^{-3}$ \\
  NS–TG           & $5.87\times10^{-1}$ & $2.23\times10^{-2}$ & $2.17\times10^{-2}$ & N/A \\
  NS–ABC          & $1.169\times10^{0}$ & $7.74\times10^{-1}$ & N/A                 & N/A \\
  \hline
  \end{tabular}
  \caption*{Emphasizing the effect of variable $i$ MOL trajectories.}
\end{table}

\begin{figure}[H]
  \centering
  \begin{minipage}[t]{0.3\textwidth}
    \centering
    \vspace{0pt}
    \includegraphics[width=\linewidth]{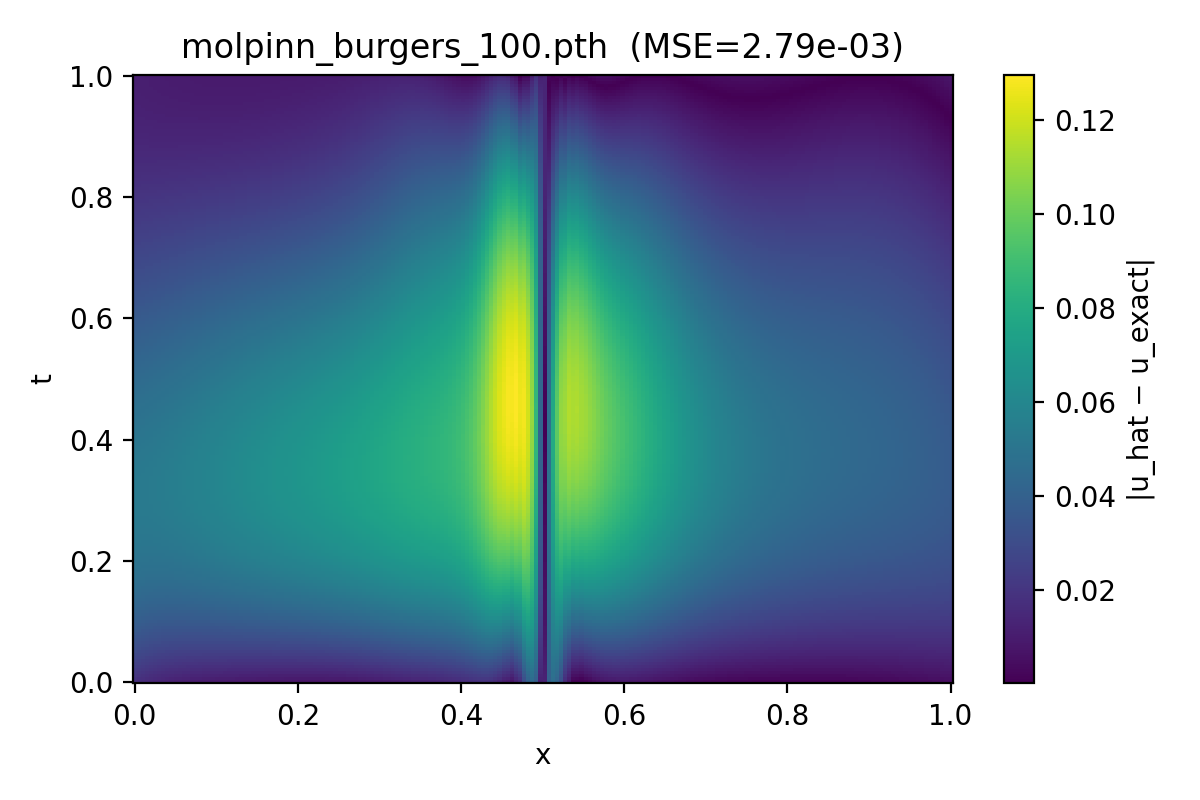}
    \caption{Image 1 caption}
  \end{minipage}
  \hfill
  \begin{minipage}[t]{0.3\textwidth}
    \centering
    \vspace{0pt}
    \includegraphics[width=\linewidth]{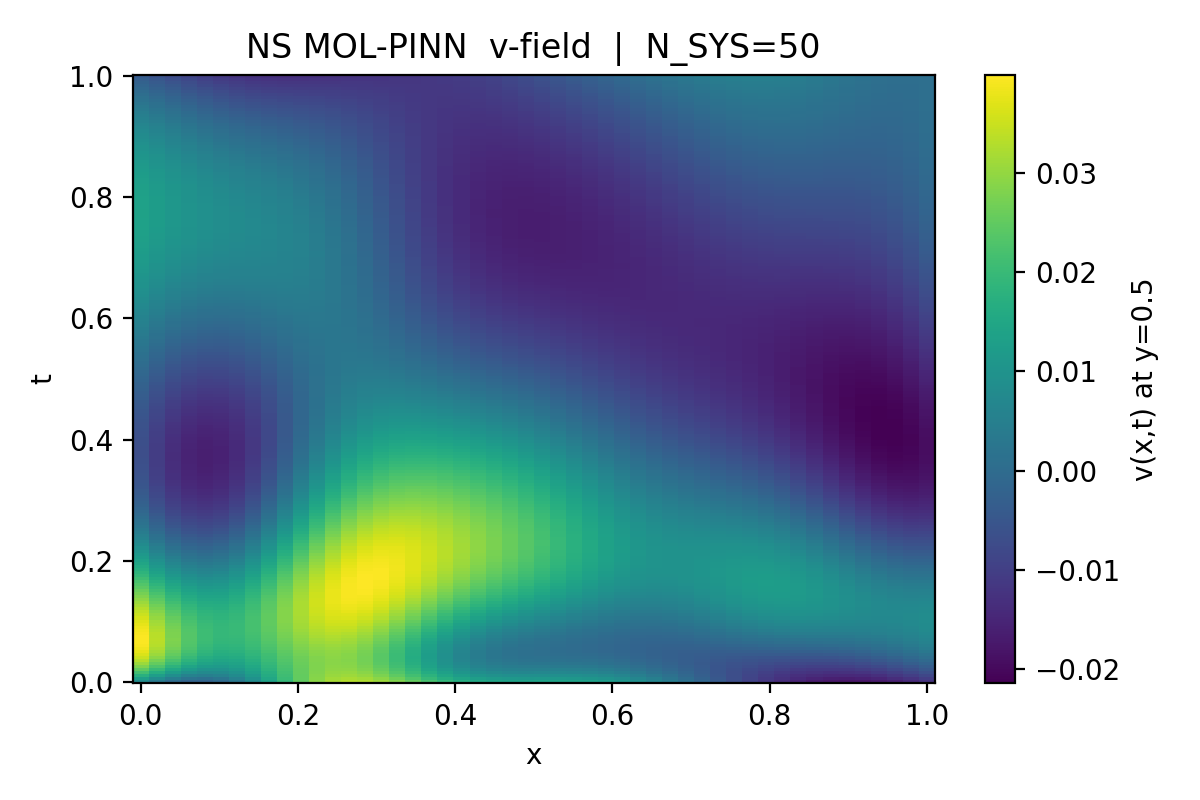}
    \caption{Image 2 caption}
  \end{minipage}
  \hfill
  \begin{minipage}[t]{0.3\textwidth}
    \centering
    \vspace{0pt}
    \includegraphics[width=\linewidth]{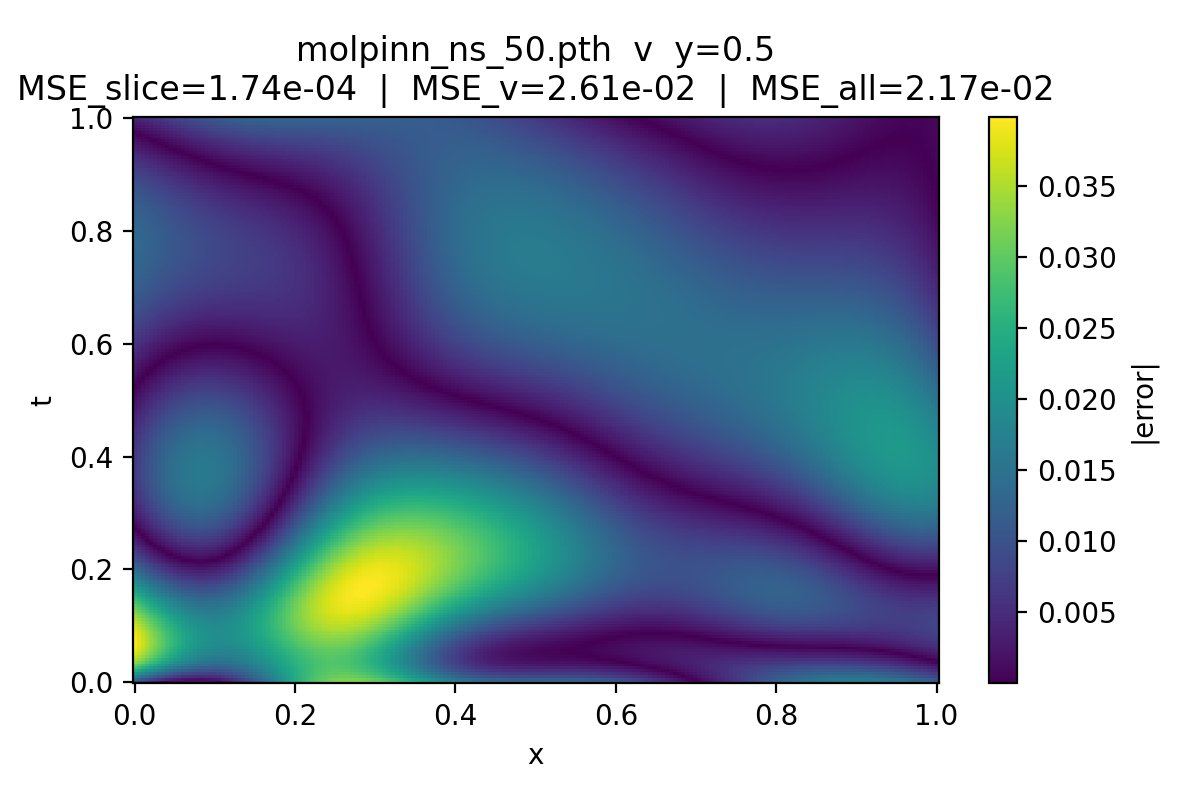}
    \caption{Image 3 caption}
  \end{minipage}
\end{figure}

\subsection{N-S Results}
\label{sec:results:N-S}
\par We evaluated MOL-PINN on N-S solutions across increasing complexity. On Taylor-Green (TG), results were markedly more consistent than PINN. Adding one dimension to test the Arnold Beltrami Childress (ABC) flow, MOL-PINN still consistently outperformed PINN and maintained similar complexity robustness. We suspect the overall low MSE reflects too few anchor functions; scaling anchors with dimensionality would likely further improve MOL-PINN over PINN.

\begin{table}[H]
  \centering
  \caption{Regional MSE for 3D/4D Navier–Stokes}
  \begin{tabular}{lccc}
  \hline
          & $v$                & $u$                & $p$ (TG) / $w$ (ABC) \\ \hline
  PINN TG         & $1.00\times10^{-2}$ & $6.58\times10^{-3}$ & $1.75\times10^{0}$  \\
  MOL-PINN TG     & $2.61\times10^{-2}$ & $2.35\times10^{-2}$ & $1.56\times10^{-2}$ \\
  PINN ABC        & $1.54\times10^{0}$  & $1.15\times10^{0}$  & $1.01\times10^{0}$ \\
  MOL-PINN ABC    & $1.29\times10^{0}$  & $6.78\times10^{-1}$ & $3.54\times10^{-1}$  \\
  \hline
  \end{tabular}
\end{table}

\subsection{Time-based Efficiency}
\label{sec:results:time}
 Table~3.1 shows that accuracy increases with more systems, but the current MOL--PINN is computationally inefficient: time per epoch also rises with system count. After 8{,}000 epochs, simulations became prohibitively long, and for this reason no data were collected for the 100-system case on Darcy and Navier-Stokes. Thus, while adding systems improves accuracy, it is accompanied by a significant computational drawback.

\begin{table}[H]
  \centering
  \caption{Time per epoch for tested simulations}
  \begin{tabular}{lcccc}
  \hline
        & \textbf{PINN} & \textbf{21i} & \textbf{50i} & \textbf{100i} \\
  \hline
  Darcy            & 0.012s     & 0.046s   & 0.142s & N/A       \\
  Burgers          & 0.0142s    & 0.017s   & 0.046s & 1.015s    \\
  Navier-Stokes 3D & 0.013s     & 0.278s   & 0.367s & N/A       \\
  Navier-Stokes 4D & 0.031s     & 2.25s    & N/A    & N/A       \\
  \hline
  \end{tabular}
\end{table}

\section{Conclusion and future work.}
\label{sec:concl}
MOL--PINN leverages automatic-differentiation gradients to achieve high-order spatial accuracy without dense meshes and shows strong resilience to solution complexity. A key limitation is an axis-aligned \textbf{directional grain bias} from the line-wise decomposition: continuity is not enforced at intermediary points, so off-grain values—--especially those far from anchored boundaries—--are harder to learn than in PINN, which can reduce accuracy on lower-complexity PDEs; runtime inefficiencies also persist at larger system counts. We plan to address these issues via two potential future frameworks: a MOL-tailored residual-based adaptive refinement pipeline to target low accuracy regions while minimizing computational necessities, and a Jacobian-enabled multi-directional MOL extension to introduce rotated grains and reduce the grain bias.

\bibliographystyle{unsrt}  


\end{document}